\documentstyle[prl,aps,multicol]{revtex}
\newcommand{\be}{\begin{equation}}
\newcommand{\ee}{\end{equation}}

\begin{document}
\draft
\title{Geometric Phase, Hannay's Angle, and an Exact Action Variable}
\author{Dae-Yup Song}
\address{Department of Physics, Sunchon National University, Sunchon 
540-742, Korea}
\date{\today}
\maketitle
\begin{abstract}
Canonical structure of a generalized time-periodic harmonic oscillator 
is studied by finding the exact action variable (invariant). 
Hannay's angle is defined if closed curves of constant action variables 
return to the same curves in phase space after a time evolution. 
The condition for the existence of Hannay's angle turns out to be 
identical to that for the existence of a complete set of (quasi)periodic 
wave functions. Hannay's angle is calculated, and it is shown that Berry's 
relation of semiclassical origin on geometric phase and Hannay's angle 
is exact for the cases considered.
\end{abstract}

\pacs{PACS number : 03.65.Bz, 03.65.Ge, 03.65.Ca}

\begin{multicols}{2}

Berry's phase \cite{Berry}, the geometric part of a change in the phase 
of a wave function under a cyclic evolution, has attracted great
interest both theoretically and experimentally \cite{SW}. 
A significant generalization of the Berry's phase, relaxing the 
adiabatic approximation, has been given by Aharonov and Anandan \cite{AA}.
The price we have to pay for this generalization is that the quantum states do 
not necessarily return to the original states up to a phase after the evolution of
Hamiltonian's cycle, so that the geometric phase may not be defined
through the method in \cite{AA}. 
This geometric phase may have a natural classical correspondent: Hannay's 
angle \cite{Hannay}.  Hannay's angle or geometric phase has long been 
studied for the (generalized) harmonic oscillator \cite{BerryH}. 
In a recent study \cite{Song} of the oscillators with time-periodic
parameters, it has been proven that, 
if two linearly independent homogeneous classical solutions of the oscillator 
are bounded for all the time, there always exists a complete set of (quasi)periodic
wave functions for an oscillator 
without driving force. For driven case, the ratio of the period of periodic 
particular solution to that of Hamiltonian should be rational for the existence of 
the complete set. 
This oscillator provides an ideal system for the study of
dynamics of Gaussian wave packets which has been applied to many problems in 
atomic and molecular physics as well as in quantum optics \cite{ASC}.

Recently, the Hannay's angle for generalized harmonic oscillators
and its relation to geometric phase were studied by Ge and Child \cite{GC}, 
and by Liu {\em et al.} \cite{Liu}. Though Hannay's angle was originally  
formulated relying on the existence of action variable in the canonical
structure of a system \cite{Hannay,BerryH}, and the {\em exact} invariant 
(action variable) for the harmonic oscillator
of time-dependent frequency has been given by Lewis \cite{Lewis} more than thirty 
years ago, none of the works for Hannay's angle of the oscillator has benefited
from the exact action variable. 

In this Letter, we will study Hannay's angle of the oscillator by generalizing
the exact action variable, and will compare the angle with the geometric phase.
The model we will consider is described by the Hamiltonian:
\begin{eqnarray}
H(x,p,t)&=& {{p}^2 \over 2 M(t)} - a(t)({p}{x}+{x}{p})
     +{1\over 2} M(t)c(t){x}^2     \cr
    && -{b(t)\over M(t)} {p}+d(t){x}
      +( {b^2(t) \over 2M(t)} -f(t)),
\end{eqnarray} 
where
\be
c(t)=w^2 + 4a^2 -2 \dot{a} -2 {\dot{M}\over M}a, ~~~
d(t)= 2ab -\dot{b} -F.
\ee
The smooth functions $M(t)$, $w(t)$ and $F(t)$ denote the positive mass, real
frequency and external force, respectively, while the overdots denote differentiations 
with respect to time. In the Lagrangian description,
the terms of smooth functions $a(t),b(t)$ and $f(t)$ can be interpreted as
the results of adding 
total derivative terms to Lagrangian which do not affect the classical equation 
of motion \cite{Song2}. We require a periodicity for every coefficient, so that
\be
H(x,p,t+\tau)=H(x,p,t).
\ee
The classical equation of motion for the Hamiltonian is given as
\be
{d \over {dt}} (M \dot{{x}}) + M(t) w^2(t) {x} =F(t).
\ee
The general solution of this equation is a linear combination of a particular
solution $x_p(t)$ and two linearly independent homogeneous solutions $u(t),v(t)$.
As in \cite{Song2}, $u(t)$ and $v(t)$ are dimensionless, while $x_p(t)$
has the dimension of length.
For later use, we define $\rho(t)$ and a time-constant $\Omega$ as
\be
\rho=\sqrt{u^2(t)+v^2(t)},
~~
\Omega = M(t)[ \dot{v}(t)u(t) - \dot{u}(t)v(t)].
\ee
The $\rho$ satisfies the following differential equation
\be
{d \over {dt}} (M \dot{\rho})-{\Omega^2 \over M\rho^3}+Mw^2 \rho=0.
\ee

For the case that $M$ is constant and $a,b,F,f$ are zero, the Hamiltonian describes
the harmonic oscillator of time-dependent frequency $w(t)$. In this case, 
the exact invariant 
\be
I_0= {1\over 2\Omega_0}[({\Omega_0x \over \rho_0})^2 + (\rho_0p-\dot{\rho}_0x)^2],
\ee
whose level surfaces form ellipses in the phase space of any time $t$,
has been found in \cite{Lewis} 
(For a derivation of $I_0$ from the Hamiltonian of a simple harmonic oscillator,
see \cite{uni}).
The subscript 0 is to denote that variables are defined 
when $M$ is the unit mass and $a,b,F,f$ are zero. The area enclosed by the ellipse is 
given by $2\pi I_0$, as can be easily proven by making use of Stokes' theorem
\begin{eqnarray}
\oint pdx &=&
\int_{I_0\geq {1\over 2\Omega_0}[({\Omega_0x \over \rho_0})^2 + 
                                         (\rho_0p-\dot{\rho}_0x)^2]} dp\wedge dx  \cr
&=&2I_0\int_{1\geq \xi_0^2+\varsigma_0^2} d \xi_0 \wedge d \varsigma_0= 2\pi I_0
\end{eqnarray}
with a parameterization $\xi_0={1\over \sqrt{2I_0\Omega_0} }(\rho_0 p-\dot{\rho}_0 x)$,
$\varsigma_0={\sqrt{\Omega_0}\over \sqrt{2I_0}}{x\over \rho_0}$. 
Though the ellipse evolves as time passes due to the time dependence of $\rho_0$, 
the area enclosed by the curve or $I_0$ is a time-constant.

To find action variable for the Hamiltonian in (1), we make use of the unitary 
transformations given in \cite{Song2}. By applying the unitary transformations 
to $I_0$, one can find that the transformed operator $I$ is written as
\begin{eqnarray} 
\label{eq:I}
I&=&{1 \over 2 \Omega}
   [{\Omega^2 \over \rho^2}(x-x_p)^2  \cr    
 & &~~~~~~~ + \{(M\dot{\rho}+2Ma\rho)(x-x_p)- \rho(p-p_p)\}^2],
\end{eqnarray}
where $p_p$ is defined as $p_p=M\dot{x}_p +2Max_p +b$. 
The generalized harmonic oscillator is a system where the path integral 
for the kernel is Gaussian. Making use of this fact, the quantum theory of the model
has been studied in Ref. \cite{Song2}. In this paper, we will consider the $I(x,p,t)$ 
in (\ref{eq:I}) as a classical object. 

One can explicitly check that $I$ is  a {\em constant of motion} satisfying
\be
{dI\over dt}= {\partial I\over \partial t} +[I, H]_{PB}=0,
\ee
where the subscript PB denotes that the term is a Poisson bracket. 
Eq. (10) shows that $I$ is the action variable of the oscillator system
described by the Hamiltonian in (1); The value of $I$ for a point of phase
space at a given time stays constant along the trajectory generated by $H$.
Again, by making use of Stokes' theorem 
as in (8), one can show that $2\pi I$ is the area of ellipse of constant $I$
in the phase space.
The term with coefficient $a$ changes the shape of ellipse, while the terms with 
coefficients $b$, $F$ just move the center of the ellipse and the purely
time-dependent term in the Hamiltonian has no effect on the ellipse.
We parameterize the ellipse as
\begin{eqnarray}
\label{eq:param}
\cos Q &=&\sqrt{\Omega \over 2I} {(x-x_p) \over \rho},\cr
\sin Q &=&{1\over \sqrt{2\Omega I}} [(M\dot{\rho}+2Ma\rho)(x-x_p)-\rho (p-p_p)].
\end{eqnarray}
Then the Poisson Bracket
\be
[I, \tan Q]_{PB}= - {d \over dQ} \tan Q
\ee
shows that $Q$ is the angle variable. 
For the evaluation of $p$ as a function of $I$ and $x$, we should consider
the two branches of an ellipse divided by the line 
$p= (M{\dot{\rho}\over \rho}+2Ma)(x-x_p) + p_p$. 
In the branch above (below) the line, the momentum is written as
\begin{eqnarray}
\label{eq:prel}
p&=&p_p +(M{\dot{\rho}\over \rho} + 2Ma)(x-x_p) \cr
&&~~~~~~
\pm {1\over \rho} \sqrt{2\Omega I -{\Omega^2 \over \rho^2}(x-x_p)^2},
\end{eqnarray}
with upper (lower) sign.
A generating function of the canonical transformation from
$\{x,p\}$ to action-angle variables is given in the branch above (below) 
the line as 
\begin{eqnarray}
\label{eq:F2}
F_2(x,I,t)&=&\delta+\int^t f(z)dz +Max_p^2+bx_p+M\dot{x}_px_p \cr
  &&+p_p(x-x_p) +({M\dot{\rho}\over 2\rho}+Ma)(x-x_p)^2 \cr 
  &&  \pm{1\over 2}\sqrt{{2\Omega I \over \rho^2}(x-x_p)^2
                 -{\Omega^2\over\rho^4}(x-x_p)^4}  \cr
 &&\mp I\tan^{-1}\sqrt{{2\rho^2I \over \Omega(x-x_p)^2} -1}
\end{eqnarray}
with upper (lower) sign. In (\ref{eq:F2}), $\delta$ is defined through the relation
\be
\dot{\delta}= {1\over 2}Mw^2x_p^2-{1\over 2}M\dot{x}_p^2.
\ee 
One can find that $p={\partial F_2\over \partial x}$ gives the relation 
(\ref{eq:prel}) and $Q={\partial F_2\over \partial I}$ is compatible with the 
parameterization (\ref{eq:param}).
The Hamiltonian for the same system in terms of action-angle variables is given as
\be
\bar{H}\equiv H(x(Q,I,t),p(Q,I,t),t)+{\partial F_2\over \partial t}
={\Omega \over M\rho^2}I,
\ee
showing again that $I$ is the action variable. 
It is noteworthy that ${\partial F_2\over \partial t}$ is
a single-valued function of $Q$ as
\begin{eqnarray}
&{\partial F_2\over \partial t}&|_{x=x(Q,I,t)}  \cr
&=& {I\over \Omega}(-M\dot{\rho}^2+{\Omega^2 \over M\rho^2} 
             -Mw^2\rho^2 +2{d(Ma)\over dt})\cos^2Q \cr
&&           + 2I{\dot{\rho}\over \rho}\cos Q\sin Q 
            +\dot{x}_p{\sqrt{2I\Omega}\over \rho} \sin Q\cr
&&+(\dot{p}_p\rho -\dot{x}_p(M\dot{\rho}+2Ma\rho))\sqrt{2I \over \Omega}\cos Q \cr
&&+x_p\dot{p}_p +\dot{\delta}+f-{d\over dt}(Max_p^2).
\end{eqnarray}

Hannay's angle is defined when the closed curves of constant action variables
return to the original curves after a time evolution \cite{Hannay,BerryH}. 
For the generalized harmonic oscillator, this can be satisfied if
both of $\rho(t)$ and $x_p(t)$ are periodic with a period which is an integral
multiple of $\tau$.  In the cases where such
period exists, we shall denote $\tau'$ as the period.  
The evolution  generated by $H(x,p,t)$  transports a family of ellipses through 
the phase space.
If $\tau'$ exists, then {\em the given family of such ellipses get transported 
back to itself after a time $\tau'$}.
Hannay's angle is defined as the integral of angle-averaged value of 
$\dot{Q}-{\partial H\over \partial I}$ for the period $\tau'$:
\be
\label{eq:def}
Q_H={1 \over 2\pi}\int_{t_0}^{t_0+\tau'}\int_0^{2\pi}{\partial \over \partial I}
     ({\partial F_2\over \partial t}|_{x=x(Q,I,t)})dQdt,
\ee
where $t_0$ is an arbitrary time.
After some algebra, one can find, making use of (6), 
that the Hannay's angle is written as:
\be
Q_H= -{1\over \Omega} \int_0^{\tau'} (M\dot{\rho}^2 +2Ma\rho\dot{\rho})dt.
\ee 

A general generating function $\tilde{F}_2(x,I,t)$ for
the action variable $I(x,p,t)$ of (\ref{eq:I}) may be
written as $\tilde{F}_2(x,I,t)= F_2(x,I,t) +IQ_c(t)+\tilde{\delta}(t)$. 
$Q_c(t)$ must be dimensionless, while $\tilde{\delta}(t)$ has the physical 
dimension of $\delta(t)$. With this general generating 
function, the angle variable $\tilde{Q}$ is given through the
parameterization of (\ref{eq:param}) where $Q$ is replaced by 
$\tilde{Q}-Q_c(t)$, so that 
\begin{eqnarray}
\label{eq:newparam}
x &=&\sqrt{2I \over \Omega} \rho \cos (\tilde{Q}-Q_c) +x_p,\cr
p &=&\sqrt{2I \over \Omega} ({M\dot{\rho} \over \rho}+2Ma) \rho \cos (\tilde{Q}-Q_c) \cr
  & &~~~-{\sqrt{2\Omega I} \over \rho}\sin (\tilde{Q}-Q_c) + p_p.
\end{eqnarray}
This parameterization is compatible with 
$\tilde{Q}= {\partial \tilde{F}_2 \over \partial I}$, and implies that 
$Q_c(t)$ represents the rotational motion of the line in phase space 
from which the angle variable is measured. 
The Hamiltonian in terms of $I$ and $\tilde{Q}$ has two additional 
terms to that of (16). Making use of the definition (\ref{eq:def}), the Hannay's angle
$\tilde{Q}_H$ for the generating function  $\tilde{F}_2$ is evaluated as  
$\tilde{Q}_H= Q_H + \Delta Q(t_0)$, where $\Delta Q(t_0)=Q_c(t_0+\tau')- Q_c(t_0)$. 
Formally, $Q_c(t)$ can be any function of $t$, and thus $\tilde{Q}_H$ depends on $t_0$ 
unless $Q_c(t)$ is periodic with the period $\tau'$.
The condition for the existence of meaningful Hannay's angle, therefore, is 
that the $Q_c(t)$ is periodic with the period $\tau'$. 
The condition is satisfied if {\em the mapping from $(x,p)$ to $(\tilde{Q},I)$
is periodic with the period $\tau'$}; If the mapping has the periodicity,
the value of $\tilde{Q}(x,p,t_0+\tau)$ is equal to that of $\tilde{Q}(x,p,t_0)$ 
in addition to the time-periodicity of $I(x,p,t)$ discussed above.
From now on we only consider the case of the periodic $Q_c(t)$, so that 
$\tilde{Q}_H$ does {\em not} depend on $Q_c(t)$ and {\em is equal to} $Q_H$. 
In (19), one can find that \cite{Song} there could be cases where the $\rho(t)$ 
and thus the $Q_H$ depend on the way of choosing $u(t)$ and $v(t)$, since the
mapping is determined by the choice of classical solutions; However, for a 
given mapping of the periodicity, the Hannay's angle is unique.   

In general cases, it looks like that
there is no reason of  Hannay's angle being described by an integral of 
canonical variables (see Ref.\cite{BerryH}).
For the generalized harmonic oscillator considered here, 
the Hannay's angle happens to be written as
\be 
\label{eq:BHdef}
Q_H=-{1 \over 2\pi}{\partial \over \partial I}\int_0^{\tau'} \int_0^{2\pi} ~ 
      p(\tilde{Q},I,t){\partial x(\tilde{Q},I,t) \over \partial t}d\tilde{Q}dt.
\ee   
The expression of Hannay's angle in (\ref{eq:BHdef})
may be facilitated in easily finding the fact that the angle does not 
depend on $x_p$ and $p_p$; 
The terms containing $x_p$ and $p_p$ in the integral of (\ref{eq:BHdef}) 
which comes from the linear terms in the Hamiltonian are
removed through the differentiation with respect to $I$ and the angle-average.
The Hannay's angle does not depend on the motion of center 
of the ellipses, reflecting the fact that the angle is a geometric quantity. 
If there is no linear term in the Hamiltonian so that $x_p=p_p=0$, 
averaging over a half range of the angle variable which corresponds to a branch 
of the ellipse 
divided by any straight line passing through origin gives the same Hannay's angle.

In the quantum treatment of the Hamiltonian in (1), a set of wave functions 
satisfying Schr\"{o}dinger equation is given as \cite{Song2}
\begin{eqnarray}
&\psi_m(x,t)&=\cr
 &&  {1\over \sqrt{2^m m!}}({\Omega \over \pi\hbar})^{1\over 4}
     {1\over \sqrt{\rho(t)}}[{u(t)-iv(t) \over \rho(t)}]^{m+{1\over 2}}
\cr&&\times
     \exp[{i\over\hbar}(\delta(t) +\int^t f(z)dz)]
\cr&&\times
     \exp[{i\over\hbar}[M(t)a(t)x^2+ (M(t)\dot{x}_p(t)+b(t))x]] 
\cr&&\times
     \exp{[{(x-x_p(t))^2\over 2\hbar}(-{\Omega \over \rho^2(t)}
               +i M(t){\dot{\rho}(t) \over \rho(t)})]}
\cr&&\times
         H_m(\sqrt{\Omega \over \hbar} {x -x_p(t) \over \rho(t)}).
\end{eqnarray}
For $\Omega > 0$, $\psi_m$ is square-integrable.
The set of wave functions $\{\psi_m(x,t)|~m=0,1,2,\cdots\}$ 
are complete with a choice of two 
linearly independent solutions $u(t), v(t)$, and a particular 
solution $x_p(t)$ \cite{Song2}, while, with the same choice of $\Omega > 0$, 
the mapping from $(x,p)$ to $(\tilde{Q},I)$ at a given time is one-to-one.

The condition for the existence of a complete set of (quasi)periodic wave 
functions is satisfied, if $\rho(t)$ and $x_p(t)$ are periodic 
with a period which is an integral multiple of $\tau$. 
Therefore, the condition of the existence of a family of time-periodic
closed curves in phase space needed for the definition of Hannay's angle is 
exactly same to that of the complete set of (quasi)periodicity. 
In \cite{Song}, it has  been shown that, if a homogeneous solution diverges
then such complete set does not exist and thus Hannay's angle can not be defined.
Making use of Floquet's theorem, it has also been proven that \cite{Song}, 
if there exist $u(t),v(t)$ finite all over the time, 
which are the cases we will consider from now on, there always 
exist two linearly independent homogeneous solutions which give periodic
$\rho(t)$ with period $\tau$ or $2\tau$; For the case of $x_p=0$,
there exists a complete set of (quasi)periodicity and thus a family 
of closed curves of periodicity. For the case of $x_p\neq 0$, 
the ratio of the period of periodic $x_p$ to $\tau$ should be rational for the 
existence of Hannay's angle and the complete set(s) of (quasi)periodicity.

If $\tau'$, an integral multiple of $\tau$, is the common period of periodic 
$\rho$ and $x_p$, the overall phase change of the wave function $\psi_m$ 
under the $\tau'$-evolution is given as \cite{Song}:
\be 
\chi_m=-(m+{1\over 2})\int_0^{\tau'}{\Omega\over M\rho^2} dt
        +{1\over \hbar}\int_0^{\tau'}(\dot{\delta}+f)dt.
\ee 
The expectation value of the Hamiltonian for a wave function $\psi_m(x,t)$ is given 
as
\begin{eqnarray}
_m<H>_m &=& \hbar(m+{1\over 2})[{\Omega\over 2M\rho^2} \cr
 &&~~~~~~  +{M\dot{\rho}^2 \over 2\Omega} + {M\rho^2 \over 2\Omega}
             -{\rho^2\over \Omega}(M\dot{a}+\dot{M}a)]     \cr
  &&+{M\over 2}\dot{x}_p^2 +{Mw^2 \over 2}x_p^2 -Fx_p  \cr
 &&  -(M\dot{a}+\dot{M}a)x_p^2 -\dot{b}x_p -f.
\end{eqnarray}
Geometric phase for the wave function $\psi_m(x,t)$ is thus written as
\begin{eqnarray}
\gamma_m &=&\chi_m + {1\over \hbar} \int_0^{\tau'}~ _m<H>_m dt \cr
        &=&(m+{1\over 2}){1\over \Omega}
             \int_0^{\tau'}[M\dot{\rho}^2+2Ma\rho\dot{\rho}]dt \cr
   &&    +{1\over\hbar} \int_0^{\tau'}
             (M\dot{x}_p^2+2Max_p\dot{x}_p+b\dot{x}_p )dt.
\end{eqnarray}

In the Hannay's angle, the effects of the linear terms in the Hamiltonian  has been
removed through the angle-average and differentiation with respect to $I$.
In the calculation of geometric phase, the expectation values of the linear terms 
can not depend on the $m$ due to the orthogonality of Hermitian polynomials and
their recurrence relations. A simple relation is thus satisfied between the geometric 
phase and the Hannay's angle
\be
Q_H = -{\partial \gamma_m \over \partial m}.
\ee
In Ref. \cite{BerryH}, the relation (26) has been suggested for general models; 
However, the suggestion was made through a semiclassical treatment of geometric 
phase and using, in fact, the Eq.(\ref{eq:BHdef}) as a defining relation for
Hannay's angle which may not be equal to the definition of Eq.(\ref{eq:def}) 
in general cases.
A similar relation for the case without linear terms is given in Ref.\cite{Liu}. 

In summary, we analyze the canonical structure of a generalized harmonic 
oscillator by finding the exact action variable. The analyses has then been used to 
find Hannay's angle which is defined relying on the time-periodic closed curves 
of constant action variables in phase space. Hannay's angle for the model considered 
can be defined if and only if there exist a complete set of
(quasi)periodic wave functions whose geometric phases have been given in this paper.
There could be cases where both of the Hannay's angle and the geometric phase 
depend on the way of choosing classical homogeneous solutions, while the angle
is unique for a given choice of classical solutions of the periodicity. 
It should be of interest if similar analyses would be possible for other models.

I like to thank Chris Jarzynski for discussions and suggestions on
the first version. Useful discussions with Peter B Gilkey are also acknowledged.

\end{multicols}


\begin{thebibliography}{99}
\bibitem{Berry} M.V. Berry, Proc. R. Soc. London Ser. A {\bf 392}, 45 (1984).
\bibitem{SW} {\it Geometric  Phases in  Physics}, edited by A. Shapere and F. Wilczek  
(World Scientific, Singapore, 1989).
\bibitem{AA} Y. Aharonov and J. Anandan, Phys. Rev. Lett. {\bf 58}, 1593 (1987).
\bibitem{Hannay} J.H. Hannay, J. Phys. A: Math.  Gen.   {\bf  18},  221  (1985).
\bibitem{BerryH} M.V. Berry,  J. Phys. A: Math.  Gen.   {\bf  18},  15  (1985);
       M.V. Berry and J.H. Hannay,  J. Phys. A: Math.  Gen. {\bf  21}, L325  (1985).
\bibitem{Song} D.-Y. Song,  Phys. Rev. A {\bf 61}, 024102 (2000).
\bibitem{ASC} {\it Quantum Optics}, edited by G.S. Agarwal and R. Inguva 
                  (Plenum, New York 1991); 
      B.L. Schumaker, Phys. Rep. {\bf 135},  317 (1986); 
      W.-M. Zhang, D.H. Feng, and R. Gilmore, Rev. Mod. Phys. {\bf 62}, 867 (1990);
      M.S. Child, {\it Semiclassical Mechanics with Molecular Applications} 
      (Oxford University Press, Oxford, 1991).
\bibitem{GC} Y.C. Ge and M.S. Child, Phys. Rev. Lett. {\bf 78}, 2507   (1997). 
\bibitem{Liu} J. Liu, B. Hu, and B. Li, Phys. Rev. Lett. {\bf 81}, 1749   (1998).
\bibitem{Lewis} H.R. Lewis, Jr.,  Phys. Rev. Lett. {\bf 18}, 510 (1967); 
        J. Math. Phys. {\bf 9}, 1976 (1968).
\bibitem{Song2}D.-Y. Song, Phys. Rev. A {\bf 59}, 2616 (1999); 
         J.   Phys.  A:   Math.  Gen.   {\bf  32},   3449  (1999).
\bibitem{uni}D.-Y. Song, Phys. Rev. A, submitted (see quant-ph/0002065). 
\end{thebibliography}
\end{document}